\documentclass{aa} 
\usepackage{txfonts}
\usepackage[latin1]{inputenc}  
\usepackage{graphicx}

\begin{document}
\title{Radiolysis of ammonia-containing ices by energetic, heavy and highly charged ions inside dense astrophysical environments}
\author{S. Pilling\inst{1,2} \and E. Seperuelo Duarte\inst{1,3,4},  E. F. da Silveira\inst{1},
E. Balanzat\inst{3},  H. Rothard\inst{3},  A. Domaracka\inst{3} \and P. Boduch\inst{3}}
 \institute{Departamento de Física, Pontifícia Universidade Católica do Rio de Janeiro (PUC-Rio), Rua Marquês de São Vicente, 225, CEP 22453-900, Rio de Janeiro, Brazil.
 \and Instituto de Pesquisa \& Desenvolvimento (IP\&D), Universidade do Vale do Paraíba (UNIVAP), Av. Shishima Hifumi, 2911, CEP 12244-000, São José dos Campos, SP, Brazil.
 \and Centre de Recherche sur les Ions, les Matériaux et la Photonique (CEA /CNRS /ENSICAEN /Université de Caen-Basse Normandie), CIMAP - CIRIL - GANIL, Boulevard Henri Becquerel, BP 5133, F-14070 Caen Cedex 05, France.
 \and Grupo de Física e Astronomia, CEFET/Química de Nilópolis, Rua Lúcio Tavares, 1052, CEP 2653-060, Nilópolis, Brazil.}
\offprints{S. Pilling,\\ \email{sergiopilling@yahoo.com.br}}
\date{Received / Accepted}
%
\abstract{ Deeply inside dense molecular clouds and protostellar disks, the interstellar ices are protected from stellar energetic UV photons. However, X-rays and energetic cosmic rays can penetrate inside these regions triggering chemical reactions, molecular dissociation and evaporation processes. We present experimental studies on the interaction of heavy, highly charged and energetic ions (46 MeV $^{58}$Ni$^{13+}$) with ammonia-containing ices H$_2$O:NH$_3$ (1:0.5) and H$_2$O:NH$_3$:CO (1:0.6:0.4) in an attempt to simulate the physical chemistry induced by heavy ion cosmic rays inside dense astrophysical environments. The measurements were performed inside a high vacuum chamber coupled to the IRRSUD (IR radiation SUD) beamline at the heavy ion accelerator GANIL (Grand Accelerateur National d'Ions Lourds) in Caen, France. The gas samples were deposited onto a polished CsI substrate previously cooled to 13 K. \textit{In-situ} analysis is performed by a Fourier transform infrared spectrometer (FTIR) at different fluences. The averaged values for the dissociation cross section of water, ammonia and carbon monoxide due to heavy cosmic ray ion analogs are $\sim 2 \times$10$^{-13}$, 1.4$\times$10$^{-13}$ and 1.9$\times$10$^{-13}$ cm$^2$, respectively. In the presence of a typical heavy cosmic ray field, the estimated half life for the studied species is 2-3 $\times 10^6$ years. The ice compaction (micropore collapse) due to heavy cosmic rays seems to be at least 3 orders of magnitude higher than the one promoted by (0.8 MeV) protons . The infrared spectra of the irradiated ice samples present lines of several new species including HNCO, N$_2$O, OCN$^-$ and NH$_4^+$ . In the case of the irradiated H$_2$O:NH$_3$:CO ice, the infrared spectrum at room temperature reveals five bands that were tentatively assigned to vibration modes of the zwitterionic glycine (NH$_3^+$CH$_2$COO$^-$).

}

\titlerunning{Radiolysis of ammonia-containing ices by heavy and energetic cosmic rays}
\authorrunning{Pilling et al.}
\maketitle

\section{Introduction}

The birthplace of stars is the densest and coldest place of the interstellar medium (ISM) called dense molecular clouds. These regions have typical gas densities of 10$^3$-10$^8$ atoms cm$^{-3}$ and temperatures of the order of 10-50 K. Due to the low temperature, dust particles (mainly silicates and carbonaceous compounds like amorphous C and SiC) can accrete molecules from the gas phase and get coated with an ice mantle. The interstellar ice mantles (or simply interstellar ices) are composed primarily of amorphous H$_2$O but usually also contain a variety of other simple molecules such as CO$_2$, CO, CH$_3$OH and NH$_3$ (e.g. Ehrenfreund \& Charnley 2000; Ehrenfreund \& Shuttle 2000; Gibb et al. 2001; Boogert et al. 2004). Laboratory studies and astronomical observations indicate that photolysis and radiolysis of such ices can create complex organic compounds, and even pre-biotic molecules like amino acids and nucleobases (e.g. Bernstein et al. 2002; Muñoz Caro et al. 2002; Kobayashi et al. 2008).

The observation of molecules in the gas phase deeply inside these cold and dense regions, where the gas sticking efficiency on grains is close to unity, suggests that they are indeed energetically active regions (e.g. Ehrenfreund \& Charnley 2000). Only a negligible amount of stellar UV photons reaches the inner parts of dense regions (Tielens \& Hagen 1982; d'Hendecourt et al. 1985). Other mechanisms have therefore been proposed to explain the presence of free molecules such as cosmic-ray-induced UV photons (Prasad \& Tarafdar 1983) and direct cosmic ray interaction with ice mantles (e.g Shen et al. 2004). Both processes lead to molecular desorption from the surface. As pointed out by Shen et al. (2004), in the case of cosmic ray particles, the sputtering promoted by direct impact and the whole grain heating (classical sublimation) due to energy deposition in the bulk are also two efficient processes to release molecules from frozen surfaces to the gas phase.

The first detection of ammonia ices has occurred around the young stellar source (YSO) NGC 7538 IRS9 (Lacy et al. 1998).  It has also been observed around the massive protostar GCS3 (Chiar et al. 2000) and several other massive YSO like W33A (Gibb et al. 2000; 2001). Ammonia has also been detected extensively inside solar system in cometary coma and on several moons (e.g. Kawakita et al. 2006; Bird et al. 1999; Moore et al. 2007 and references therein).

One of the most reactive and important molecules observed from photolysis and radiolysis experiments of ammonia-containing ices, in which there is a carbon source, is the cyanate ion, OCN$^-$ (e.g. Demyk et al. 1998; Hudson \& Moore 2000; van Broekhuizen et al. 2005 and references therein). Its infrared band around 2165 cm$^{-1}$ was observed in several protostellar sources (e.g. Lacy et al. 1984; Gibb et al. 2000; Whittet et al. 2001) indicating regions were UV photochemistry and ion/electron bombardments have played an important role in the chemical alteration of grain mantles.

Although the flux of heavy ions (e.g. Fe, Ni, Si, Mg, ...) is about 3-4 orders of magnitude lower than that of protons (Roberts et al. 2007; Mennella et al. 2003), their effects play an important role on interstellar grains since they can deposit  about 100 times more energy than the light ions (He$^+$ and protons) inside the grains. Brown et al. (1984) and Fama et al. (2008) using high energy He$^+$ and protons have found that the sputtering yields scales with the square of the electronic stopping power. We have recently proved that this square law extends its validity up to values of electronic stopping, such as those induced by swift heavy ions (Seperuelo Duarte et al. 2009a,b). Consequently, amount of species released per impact to gas phase due to heavy ions could be 4-5 orders of magnitude higher than those for protons.

In order to characterize the effects of heavy ions on different interstellar ice analogs, it is necessary to measure their sputtering yields and chemical reaction cross sections and to compare them to those promoted by He$^+$ and protons. Studies on the effect of heavy projectiles in astrophysical ice analogs are scarce. Most of experiments have been performed in the keV or hundred of keV range, mainly with protons, He and Ar ions (e.g. Loeffler et al. 2006; Gomis et al. 2004a,b). Due to the low kinetic energy of these ions, the structural and chemical changes in the frozen ices are mainly governed by nuclear energy loss in elastic ion-target atom collisions (e.g. Brown et al. 1982; 1984).

In this work, we present infrared measurements of ammonia-containing astrophysical ice analogs (H$_2$O:NH$_3$ (1:0.5) and H$_2$O:NH$_3$:CO (1:0.6:0.4)) irradiated by 46 MeV $^{58}$Ni$^{13+}$ to simulate the modification induced by heavy cosmic rays inside dense astrophysical environments. At such high ion velocity, the energy deposition  is mainly due to inelastic electronic interactions with target electrons (electronic stopping power regime).

\section{Experimental}

\begin{figure}[!t]
 \centering
 \includegraphics[scale=0.75]{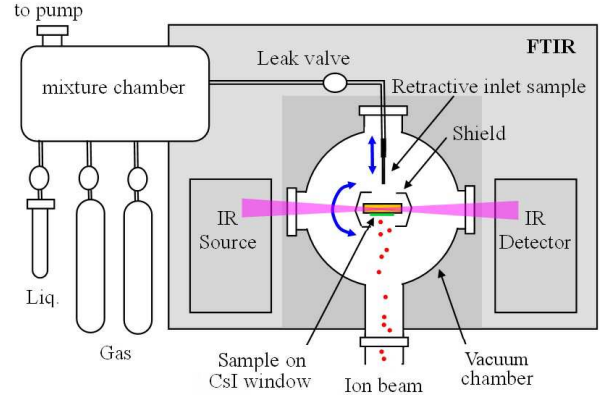}
\caption{Schematic diagram of the experimental set-up. The ion beam impinges perpendicularly on the thin ice film deposited on a CsI crystal.} \label{fig:diagram}
\end{figure}

We have used the facilities at the heavy ion accelerator GANIL (Grand Accelerateur National d´Ions Lourds) in Caen, France.  A 46 MeV $^{58}$Ni$^{13+}$ ion projectiles impinge perpendicularly on the ice target. In this paper, results for two ammonia-containing ices H$_2$O:NH$_3$:CO (1:0.5:0.4) and H$_2$O:NH$_3$ (1:0.5) are presented. The incoming charge state $13+$ corresponds approximately to the equilibrium charge state after several collisions of a 46 MeV Ni atoms (independent of the initial charge state) with matter (e.g. Nastasi et al. 1996).

\emph{In-situ} Fourier transformed infrared (FTIR) spectra were recorded for ices irradiated at different fluences, up to  $2 \times 10^{13}$ ions cm$^{-2}$ using a Nicolet FTIR spectrometer (Magna 550) from 4000 to 650 cm$^{-1}$ with 1 cm$^{-1}$ resolution. The ion flux was $2 \times 10^{9}$ cm$^{-2}$ s$^{-1}$. A background allowing the absorbance measurements is collected before gas deposition. The total irradiation time was about 3 hours for each sample. For further experimental details see Seperuelo Duarte et al. (2009a).

The sample-cryostat system can be turned over 180$^\circ$ and fixed in three different positions to allow: i) gas deposition, ii) FTIR measurement and iii) perpendicular irradiation as shown in Fig~\ref{fig:diagram}. The thin ice film was prepared by condensation of gases (purity better than 99\%) onto a CsI substrate attached to a closed-cycle helium cryostat,  cooled to 12-13 K. A very small fraction of the mixture CO$_2$, CO$^{18}$O and C$^{18}$O$_2$ existed as contaminants in the pre-chamber. During the experiment the chamber pressure was around $ 2\times 10^{-8}$ mbar.

Following d´Hendecourt \& Allamandola (1986) the molecular column density of a sample was determined from the relation between optical depth $\tau_\nu = \ln(I_0/I)$ and the band strength, A (cm molec$^{-1}$), of the sample respective vibrational mode. In this expression $I$ and  $I_0$ is the intensity of light at a specific frequency before and after passing through a sample, respectively. Since the absorbance measured by FTIR spectrometer was $Abs_\nu=\log(I_0/I)$ the molecular column density of ice samples was given by:
\begin{equation} \label{eq-N}
N= \frac{1}{A} \int \tau_\nu d\nu = \frac{2.3}{A} \int Abs_\nu d\nu \quad \textrm{[molec cm$^{-2}$]}
\end{equation}
where $Abs_\nu = \ln(I_0/I)/\ln(10) = \tau_\nu/2.3$.

From these measurements and assuming an average density for the ice samples of about 1 g/cm$^3$ (water ice), the thickness and the deposition rate are determined. For the H$_2$O:NH$_3$:CO ice (1:0.6:0.4) mixture the thickness was of about 1.7 $\mu$m and the deposition rate of $\sim$ 13 $\mu$m/h. For the H$_2$O:NH$_3$ ice (1:0.5) the thickness was $\sim$ 1.4 $\mu$m and the deposition rate was about 10.5 $\mu$m/h.

The analyzed ice layers were thin enough: i) to avoid saturation of the FTIR signal in transmission mode and ii) to be fully crossed by ion beam with approximately the same velocity. This latter point is important since a relatively small total kinetic energy loss of the projectile in the film guarantees that the studied cross sections remain constant.

The vibrational band positions and their infrared absorption coefficients (band strengths) for used in this work are given in Table~\ref{tab:A}.

\begin{table}[!tb]
\caption{Infrared absorption coefficients (band strengths) used in the column density calculations for the observed molecules.} \label{tab:A}
\setlength{\tabcolsep}{4pt}
\begin{tabular}{ l l l l r }
\hline \hline
Frequency  & Wavelength & Assignment  & Band strength (A)           & Ref. \\
(cm$^{-1}$) & ($\mu$m)  &             &   (cm molec$^{-1}$)      &      \\
\hline
2342         & 4.27  & CO$_2^a$ ($\nu_3$)   & 7.6$\times 10^{-17}$   & [1]  \\
$\sim$ 2234  & 4.48 & N$_2$O ($\nu_3$)     & 5.2$\times 10^{-17}$   & [2]  \\
$\sim$ 2165  & 4.62 & OCN$^-$ ($\nu_3$)    & 4$\times 10^{-17}$     & [3]  \\
2139         & 4.67 & CO ($\nu_1$)         & 1.1$\times 10^{-17}$   & [4]  \\
$\sim$ 1100  & 9.09 & NH$_3$ ($\nu_2$)     & 1.2$\times 10^{-17}$   & [5]  \\
$\sim$ 800   & 12.5 & H$_2$O ($\nu_L$)     & 2.8$\times 10^{-17}$   & [4]  \\
\hline \hline
\multicolumn{4}{l}{$^a$ CO$^{18}$O and C$^{18}$O$_2$ at 2325 and 2309 cm$^{-1}$, respectively.}\\
\end{tabular}
[1] Gerakines et al. 1995; [2] Wang et al. 2001; [3] d'Hendecourt \& Allamandola 1986; [4] Gibb et al. 2004; [5] Kerkhof et al. 1999.
\end{table}

\section{Results}

\begin{figure}[!t]
 \centering
 \includegraphics[scale=0.72]{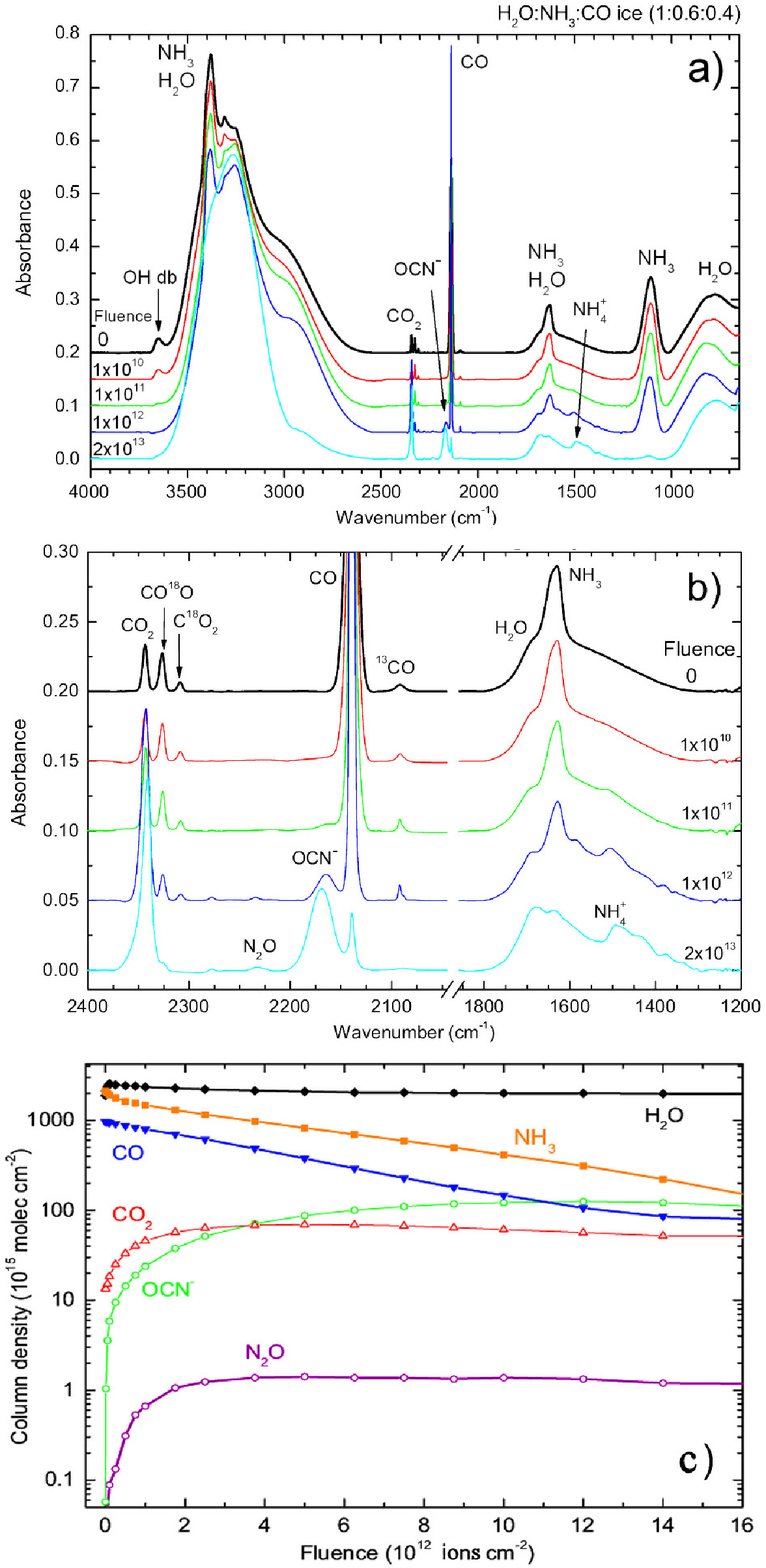}
\caption{(a) Infrared spectra of H$_2$O:NH$_3$:CO ice (1:0.6:0.4) before (top dark line) and after different irradiation fluences. (b) Expanded view from 2400 to 1200 cm$^{-1}$. Each spectrum has an offset of 0.05 for better visualization. (c) Molecular column density derived from the infrared spectra during the experiment.} \label{fig:EXP1-FTIR}
\end{figure}

Figure~\ref{fig:EXP1-FTIR}a-b presents the infrared spectra of H$_2$O:NH$_3$:CO ice (1:0.6:0.4) before (highest curve) and after different irradiation fluences.  Each spectrum has an offset of 0.05 for better visualization. The narrow peak at 2100 cm$^{-1}$ is the CO  stretching mode ($\nu_1$). The broad structure from 3100 to 3500 cm$^{-1}$ presents a combination of between vibration modes of water ($\nu_1$) and ammonia ($\nu_1$). The band at 1600 cm$^{-1}$ is composed by two lines corresponding to the water $\nu_2$ vibration mode (1650 cm$^{-1}$) and ammonia $\nu_4$ vibration mode (1630 cm$^{-1}$).  The feature around 1100  cm$^{-1}$ is the umbrella vibration ($\nu_2$) mode of ammonia and the one at 800 cm$^{-1}$ is the libration mode ($\nu_L$) of water molecules. The OH dangling bond (OH db) line at about 3650 cm$^{-1}$ is also observed. This band indicates a high degree of porosity (Palumbo 2006) in this ice mixture and will be discussed later. The region between 2400 to 1200 cm$^{-1}$ is shown in detail in Figure~\ref{fig:EXP1-FTIR}b. We can see the ($\nu_3$) band of newly formed cyanate ion OCN$^-$ at 2160 cm$^{-1}$ and NH$_4^+$ around 1500 cm$^{-1}$. Increasing of CO$_2$ abundance as a function of fluence is mainly due to following processes (CR symbolized cosmic rays):
\begin{equation}
\textrm{CO} \stackrel{CR}{\longrightarrow} \textrm{C} + \textrm{O} \\
\end{equation}
\begin{equation}
\textrm{CO} + \textrm{O} \rightarrow \textrm{CO}_2 \\
\end{equation}

The variation of the column density of the most abundant molecules observed during the irradiation of H$_2$O:NH$_3$:CO ice (1:0.6:0.4) by 46 MeV Ni ions is shown in Fig.~\ref{fig:EXP1-FTIR}c as a function of fluence. The column density of water was determined taking into an account the integrated value of IR absorbance between 1025 cm$^{-1}$ and 800 cm$^{-1}$ (approximately the half left part the IR libration band) and multiplying it by a factor 2. Since this procedure was done for all IR spectra, the relative error is expected to be small. In the case of CO$_2$, the column density is the sum of all its isotopic contributions. For both CO$^{18}$O and C$^{18}$O$_2$ molecules, the same value of band strength determined for the $\nu_3$ vibration mode CO$_2$ at 2342 cm$^{-1}$ (Gerakines et al. 1995) was adopted.

The ammonia and CO abundances seem to present a similar dissociation rate reaching the half of initial values at a fluence of about 4 $\times 10^{12}$ ions cm$^{-2}$. The column density of water is constant, about 2 $\times 10^{18}$ molecules cm$^{-2}$, decreasing very slowly as the fluence increases. This is attributed to a persistent deposition of water from the residual gas. This effect was not observed in our previous experiments on pure CO and CO$_2$ ices (Seperuelo Duarte et al. 2009a,b) and is then apparently related to the deposition of water containing ices.)

The OCN$^-$ abundance increases very fast, reaching a maximum at a fluence between 1.2 and 1.3 $\times 10^{13}$ ions cm$^{-2}$. The N$_2$O production reaches a maximum at around 5 $\times 10^{12}$ ions cm$^{-2}$ and remains constant until the end of irradiation (up to a fluence of 2 $\times 10^{13}$ ions cm$^{-2}$). As pointed out by Jamieson et al. (2005), N$_2$O was observed in the radiolysis of N$_2$-CO$_2$ ices. They have proposed a pathway reaction between N$_2$ and the oxygen atom coming from the dissociation of CO$_2$ by 5 keV electrons. N$_2$O was also reported in both N$_2$+CO and N$_2$+CO$_2$ ice irradiations by 0.8 MeV protons from a Van de Graff accelerator performed by Hudson \& Moore (2002). Recently, we performed additional experiments on N$_2$-CO (1:1) and NH$_3$-CO (1:1) ices irradiated by 537 MeV Ni projectiles. N$_2$O  species have been observed for the N$_2$-CO system, confirming the results reported by Jamieson et al. (2005) and by Hudson \& Moore (2002), but not for NH$_3$-CO. The conclusion is that N$_2$ molecules are not easily formed from NH$_3$ and that they are required as an intermediate step to synthesize N$_2$O.

\section{Discussion}

\subsection{Dissociation cross section}

\begin{table}[!b]
\begin{center} %
\caption{Stopping power and penetration depth values for equivelocity 46 MeV $^{58}$Ni and 0.8 MeV protons in water ice calculated by SRIM code. } \label{tab:dedxSRIM}
\setlength{\tabcolsep}{3.5pt}
\begin{tabular}{ l l l c  }
\hline \hline
ion    & \multicolumn{2}{c}{Stopping Power}  & Penetration depth  \\
        & \multicolumn{2}{c}{(keV$/\mu$m)}    & ($\mu$m)         \\
        \cline{2-3}
        & electronic      & nuclear    &                      \\
\hline
0.8 MeV p       & 29     & 0.02            & 17.7 \\
46 MeV Ni    & 4900   & 15              & 19.5 \\
\hline \hline
\end{tabular}
\end{center}
\end{table}  %

In the present case of fast ions (projectile velocity $\gtrsim$ 0.2 cm ns$^{-1}$), most of the deposited energy leads to excitation/ionization of target electrons. In turn, the electrons liberated directly from the inner part of the ion track, about 0.3 nm of diameter for 46 MeV Ni ions (Iza et al. 2006), transfer most of their energy to the surrounding condensed molecules ($\sim$ 3 nm). The re-neutralization of the track proceeds concomitantly with the local temperature rise, leading to an eventual sublimation. At the ice surface, the energy delivered by these heavy ions allows molecules to be removed by sputtering, causing the formation of craters of a few nm deep (Schmidt et al. 1991). The sputtering yield, $Y$, scales with the square of the electronic stopping power inside the target, $S_e = $d$E/$d$x$   (e.g. Brown et al. 1984; Famá et al. 2008, Seperuelo Duarte et al. 2009a,b).

\begin{table*}[!t]
\begin{center} %
\caption{Dissociation cross sections of the studied molecular species for radiolysis of ammonia-containing ices by 46 MeV Ni.} \label{tab:SIG-Destruct}
\setlength{\tabcolsep}{4pt}
\begin{tabular}{ l l c c c c c c }
\hline \hline
Species  & Mixture & $\sigma_d$  & $N_\infty$  & $Y$   &  $L^f$ &  $N_0$   & $Model$    \\
 & (H$_2$O:NH$_3$:CO) & $(10^{-13}$ cm$^2$)  & (10$^{17}$ molec cm$^{-2}$)  & (10$^4$ molec ion$^{-1}$)  & (10$^4$ molec ion$^{-1}$) & (10$^{17}$ molec cm$^{-2}$)   &    \\
\hline
H$_2$O   & (1:0.5:0)     & $\sim 2$  & 23        & 1$^a$    & 38    & 29      & 1 \\
         &  (1:0.6:0.4)  & $\sim 2$  & 19        & 1$^a$    & 35    & 24      & 2 \\
NH$_3$   & (1:0.5:0)     & 1.3       & NA$^e$    & 0$^b$    & 0      & 2.0     & 3 \\
         & (1:0.6:0.4)   & 1.4       & NA        & 0$^b$    & 0      & 1.7     & 4 \\
CO       & (1:0.6:0.4)   & 1.9       & NA        & 0$^b$    & 0      & 1.0     & 5a \\
         & (1:0.6:0.4)   & 1.9       & NA        & 0$^b$    & 1$^d$  & 1.0     & 5b \\
         & (1:0.6:0.4)   & 1.9       & NA        & 1$^c$    & 0      & 1.0     & 5c\\
\hline \hline
\end{tabular}
\end{center}
$^a$ Taken from Brown et al. (1984).
$^b$ No sputtering. Assuming that the water layering is thick enough to fully cover the NH$_3$ or CO molecules on the surface.
$^c$ Assuming that there is no water layering and considering a sputtering of 1$\times 10^4$ molec ion$^{-1}$ (the same adopted for water molecules).
$^d$ Assuming an extra source of CO for Layering.
$^e$ NA=Not applied.
$^f$ For H$_2$O: derived from $L = N_\infty \sigma_{d}+Y$. For NH$_3$ and CO: assuming no layering (except model 5b).
\end{table*}

The electronic and nuclear stopping power values for 46 MeV $^{58}$Ni and its equivelocity protons (0.8 MeV protons) in water ice ($\rho = 1$ g cm$^{-3}$), calculated by SRIM\footnote{www.srim.org} - Stopping and Ranges of Ions in Matter code (Ziegler \& Biersack, 2006), are given in Table~\ref{tab:dedxSRIM}. Their penetration depths in the water ice are also shown. The electronic stopping power is at least 2 orders of magnitude higher than the nuclear one; the total stopping power for 46 MeV Ni atoms is about 2 orders of magnitude higher than that of equivelocity protons.

The variation in the molecular abundance due to the incoming radiation can be attributed to several processes: i) the molecular dissociation quantified by the dissociation cross section, $\sigma_{d}$ (cm$^{2}$); ii) the sputtering yield, $Y$ (molecules desorbed per ion impact); iii) the formation of molecules, $\sigma_{f}$ (cm$^{2}$), from species already present as well as from radicals and ionic fragments; iv) the molecular layering, $L$ (molecules deposited per ion impact), due to deposition of the residual gas or the eventual recondensation of sputtered molecules. The layering yield, $L$, is expressed in terms of molecules deposited by ion impact; it is obtained by dividing the layering rate (molec s$^{-1}$) by the ion projectile flux ($ 2 \times 10^9$ ions s$^{-1}$).

Taking into account the two ``formation" processes (chemical reaction and layering), as well as the two ``disappearance" processes (molecular dissociation and sputtering, the column density rate for each molecular species is given by:
\begin{equation} \label{eq:PAI}
\frac{dN_i}{dF} = \sum_{j \neq i} \sigma_{f,ij} N_j + L_i  - \sigma_{d,i} N_i - Y_i \Omega_i(F)
\end{equation}
where  $\sum_j \sigma_{f,ij} N_j$ represents the total molecular production rate of the $i$ species directly from the $j$ species, $Y_i$ is the sputtering yield of a pure ice formed by $i$ species, and $\Omega_i(F)$ is the relative area occupied by the $i$ species on the ice surface.

Assuming: i) that only $i$ = 1 species (e.g. water) of the residual gas are able to condense on the sample (i.e., $L_i = L_1 \delta_{i1}$, where $\delta_{ij}$ is the Dirac function), and ii) that the analyzed molecular species cannot react in a one-step process to form another species originally present in the ice (i.e., $\sigma_{f,ij} \approx 0$), this equation system becomes:

\begin{equation} \label{eq:enio3}
\frac{dN_i}{dF} = - \sigma_{d,i} N_i - (Y_i  \Omega_i(F) - L_1 \delta_{i1})
\end{equation}

\begin{figure}[!t]
 \centering
\includegraphics[scale=0.77]{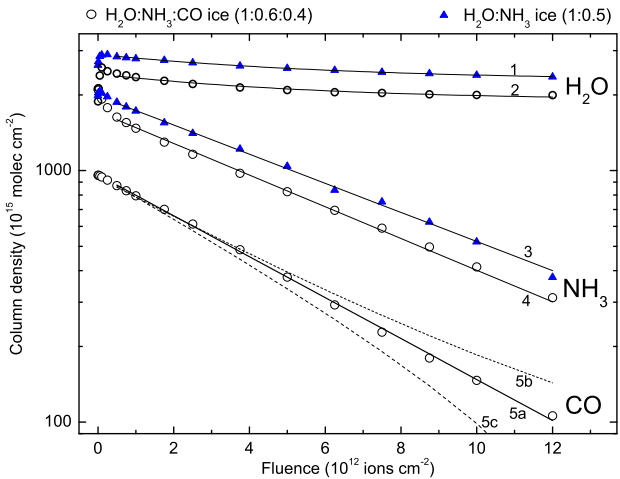}
\caption{Variation of the column density of water, ammonia and CO as a function of fluence. Solid lines are the fittings using Eq.~\ref{eq:enio4} and Eq.~\ref{eq:enio5}. Triangles and circles correspond to different ice samples. Model parameters are listed in Table~\ref{tab:SIG-Destruct}.} \label{fig:fitting}
\end{figure}

As shown in Fig.~\ref{fig:fitting} and Fig.~\ref{fig:zoom-compact}a (hachured region), there is a non-exponential decreasing behavior at the beginning of the irradiation, which boundary is defined by $F < F_{min} \approx 5 \times 10^{11}$ ions cm$^{-2}$. This suggests that ice compaction is occurring, as it will discussed further.

Another phenomenon that can also occur is layering of species 1 over the ice surface ($\Omega_i(F) \rightarrow \delta_{i1}$), preventing progressively sputtering of $i\neq 1$ species. At the end of these processes, the equation system (\ref{eq:enio3}) is decoupled and can be solved analytically:

\begin{equation} \label{eq:enio4}
N_i = N_{0i} \exp(-\sigma_{d,i} F)  \quad \textrm{, for $i\neq 1$ species}
\end{equation}
and

\begin{equation}\label{eq:enio5}
N_1= (N_{01} - N_{\infty,1}) \exp(-\sigma_{d,1} F) + N_{\infty,1}
\end{equation}
where $N_{\infty,1} = (L_1 - Y_1)/ \sigma_{d,1}$ is the asymptotic value of column density at higher fluences due to the presence of layering.
$N_i$ and $N_{0i}$ are the column density of species $i$ at a given fluence and at the beginning of the experiment, respectively. Eq.~\ref{eq:enio5} describes the condensation of $i=1$ species during irradiation. This equation is particulary useful for species which adsorbs on the walls of vacuum chamber as the case of water, generating a continuous flow towards the cold target, creating on it a superficial layer during the experiment.

Fig.~\ref{fig:fitting} presents the best fittings of the infrared data of ices H$_2$O:NH$_3$:CO and H$_2$O:NH$_3$, employing equations~\ref{eq:enio4} and \ref{eq:enio5} for describing the column density evolution with the fluence. The condition $\sigma_{f,1j} \approx 0$
applied for water means that the production rate of water molecules from ammonia and carbon monoxide is negligible compared to the layering rate.
Since $F_{min} \approx 5 \times 10^{11}$ ions cm$^{-2}$ for the current data, the fittings cover only the $F = 1-12 \times 10^{12}$ ions cm$^{-2}$ range to avoid both the mentioned transient effects and eventual large inhomogeneities in the ice induced at too large fluences. The observed decreasing of $N_i$ for CO and NH$_3$ ices is indeed exponential, leading to the CO dissociation cross section the value of $1.9 \times 10^{-13}$ cm$^2$, in agreement with the value obtained for pure CO ice irradiated by the same projectiles (Seperuelo Duarte et al. 2009b). The values for NH$_3$ are 1.3 and 1.4 $\times 10^{-13}$ cm$^2$, obtained with the H$_2$O:NH$_3$ and H$_2$O:NH$_3$:CO ice, respectively. The fitting parameters (dissociation cross section, sputtering, layering and the initial relative molecular abundance of each species in the ices) are listed in Table~\ref{tab:SIG-Destruct}. The individual effects of layering or sputtering on the column density (e.g. for CO species) correspond to the models 5b and 5c, respectively.

The H$_2$O column density data, presented in Fig.~\ref{fig:EXP1-FTIR}c, level off around $2 \times 10^{18}$ molec cm$^{-2}$ for the two samples. The average value for the dissociation cross section of water, employing Eq.\ref{eq:enio4}, is $\sigma_{d,1} \sim 2 \times 10^{-13}$ cm$^2$. The sputtering yield for water measured by Brown et al. (1984) was extrapolated for the 46 MeV Ni ions impact by using the stopping power displayed on Table~\ref{tab:dedxSRIM}. The obtained value is $Y_1 = 1 \times 10^4$ molecules per impact; the estimated average water layering of both experiments is $4 \times 10^{5}$ molec ion$^{-1}$, obtained from the relation $L = N_\infty \sigma_{d}+Y$. Due to the large value of water layering, the NH$_3$ and CO species were recovered by a H$_2$O film during irradiation and their sputtering yields were considered negligible; therefore, the data were adjusted directly by Eq.\ref{eq:enio4}.

\subsection{Ice compaction}

\begin{figure}[!tb]
\centering
\includegraphics[scale=0.38]{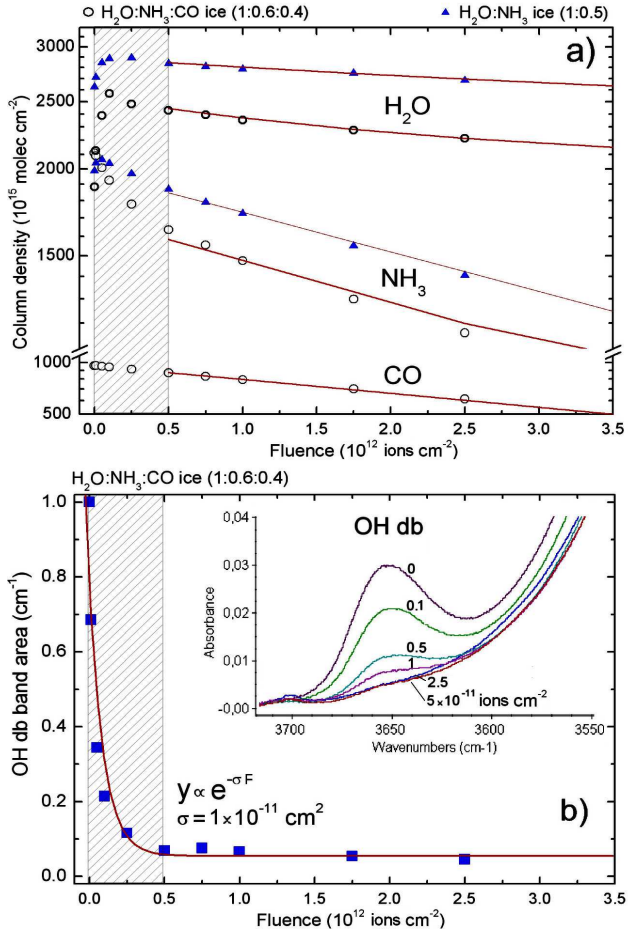}
\caption{(a) Expanded view of Fig.~\ref{fig:fitting} at the beginning of irradiation. Note the drastic changes in column density of water and ammonia up to fluence of 5$\times 10^{11}$ ions cm$^{-2}$. Hachured strip indicates the regions where the compaction is occurring. (b) OH dangling bond feature area (cm$^{-1}$) in the H$_2$O:NH$_3$:CO ice (1:0.6:0.4) as a function of fluence. Figure inset is the OH db feature from non irradiated ice and at fluences 0.1, 0.5, 1, 2.5 and 5$\times10^{11}$ ions cm$^{-2}$.} \label{fig:zoom-compact}
\end{figure}

A high density amorphous form of water ice is found when water vapor is deposited on very cold (T $<$ 30 K) surfaces, at a rate $<$ 100 $\mu$m/h (Jenniskens, Blake \& Kouchi 1998). Both of our experiments are within this regime, nevertheless the ice mixture containing CO was found to be less compact than the H$_2$O:NH$_3$ ice. This conclusion is consistent with the observation of large changes in the column density of the adsorbed species (Fig.~\ref{fig:zoom-compact}a) and is supported by the presence of the OH dangling bond (OH db) line around 3650 cm$^{-1}$ (Fig.~\ref{fig:zoom-compact}b), which is attributed to water molecules in the surfaces of micropores (Rowland \& Devlin 1991; Rowland et al. 1991; Palumbo 2006 and references therein). Since both ices were deposited roughly at the same rate ($\sim$ 12 $\mu$m/h) and at the same temperature ($\sim$ 13 K), the porosity enhancement must be attributed to the CO molecules, which diffuse very efficiently into the ice (e.g. Collings et al. 2004). Some cavities are eventually created during the ice formation reducing its compaction.

Figure~\ref{fig:zoom-compact}a presents an expanded view of Figure~\ref{fig:fitting} corresponding to fluences up to $3.5 \times 10^{12}$ ions cm$^{-2}$, i.e. at the beginning of the irradiation. The hatched region ($F < F_{min}$) indicates the fluences for which the column densities present a complex behavior. Such rapid variations, observed for both water and ammonia species, are enhanced in the case of H$_2$O:NH$_3$:CO mixture but are not seen in the CO column density. This fact may be associated with changes in the band strength of vibration modes of hydrogen-bound molecules produced by compaction effects at the beginning of the irradiation.

A direct evidence of the compaction (micropore collapse) due to the impinging heavy ions is shown in Figure~\ref{fig:zoom-compact}b, where the intensity of the OH db band as function of fluence is examined. The experimental data (band area, $y$) have been fitted by the exponential curve $y \propto e^{-\sigma F}$ where $F$ is the ion fluence (ions cm$^{-2}$) and $\sigma = 1 \times 10^{-11}$ cm$^2$ is the disappearance cross section. The estimated error is within 30\%. The figure inset presents the OH db feature at 3650 cm$^{-1}$ during the beginning of the irradiation. The traces correspond to fluences 0 (non-irradiated), 0.1, 0.5, 1, 2.5 and 5$\times10^{11}$ ions cm$^{-2}$. The peculiar changes in the column density, at the beginning of the irradiation, and the fact that the OH db feature seems to disappear approximately at the same fluence, around 5$\times 10^{11}$ ions cm$^{-2}$, indicate that at this point the ice is fully compacted by the Ni ion beam.

A detailed investigation on the compaction effects due to ion bombardment was performed by Palumbo (2006) using 200 keV protons in water-rich ices at 15 K.  The intensity of the OH db line decreases after ion irradiation and the amount of absorbed carbon monoxide (from an upper layer of CO ice) also decreases as the fluence of impinging ions increases. These results indicate that the porosity of amorphous water ice decreases after ion irradiation. Palumbo (2006) has quantified the compaction degree of water ice irradiated by 200 keV protons by fitting an exponential function to the integrated area of the OH db peak at different fluences obtaining a ``disappearing" cross section of 4.13 $\times 10^{-14}$ cm$^{2}$. Comparing the cross section of both experiments, it is observed that compaction effects due to heavy and energetic ions (46 MeV Ni ions) are 3 orders of magnitude higher than the one promoted by (0.8 MeV) protons.

\subsection{Synthesis of molecules}

\subsubsection{Cyanate ion, OCN$^-$}

The 2165 cm$^{-1}$ (4.62 $\mu$m) peak, commonly referred to the ``XCN" band, was first detected toward the massive protostar W33A (Soifer et al. 1979). Its presence was the first observational indication that complex chemistry (triggered by UV/X-ray photons or ion/electron bombardments) could be occurring in the interstellar ice mantles. To date this feature has already been observed in more than 30 deeply embedded mostly high-mass young stellar objects (Tegler et al. 1993; Demyk et al. 1998; Pendleton et al. 1999; Gibb et al. 2000; Keane et al. 2001; van Broekhuizen et al. 2005; Pontoppidan et al. 2003; Whittet et al. 2001; Gibb et al. 2004) and several galactic center sources (Chiar et al. 2002; Spoon et al. 2003).

The ``XCN" band has been studied extensively in the laboratory, being easily produced by proton irradiation (e.g. Moore et al. 1983; Hudson et al. 2001), vacuum ultraviolet photolysis (e.g. van Broekhuizen et al. 2005; Lacy et al. 1984; Grim \& Greenberb 1987; Demyk et al. 1998), or thermal annealing of interstellar ices analogues (Raunier et al. 2003; van Broekhuizen et al. 2004). Along the years, several candidates have been proposed to justify the XCN band, including some isonitriles (X-N$\equiv$C) and cyanates (X-O-C$\equiv$N) (see Pendleton et al. 1999 for review). However, laboratory experiments involving isotopic substitution proved unequivocally that the responsible for the 2165 cm$^{-1}$ feature is the cyanate ion, OCN$^-$ (Schutte \& Greenberg 1997; Bernstein et al. 2000; Novozamsky et al. 2001; Palumbo et al. 2000).

The present work is the first report on the OCN$^-$ formation by heavy, highly charged and energetic ions (46 MeV $^{58}$Ni$^{13+}$), as an attempt to simulate the chemistry and the physical chemistry induced by cosmic rays inside dense regions of interstellar medium like dense molecular clouds or protoplanetary disks.

\begin{figure}[!t]
\centering
\includegraphics[scale=0.74]{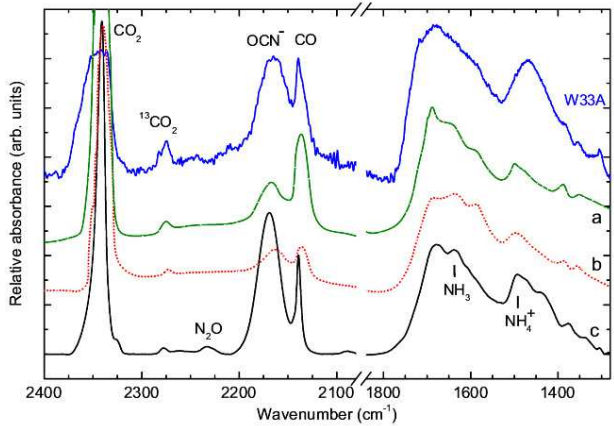}
\caption{Comparison between IR spectra of interstellar and laboratory ices. The highest curve is the infrared spectra of protostellar source W33A obtained by Infrared Space Observatory (ISO). Lower traces indicate lab spectra of H$_2$O:NH$_3$:CO ices after processing by: (a) UV photons (Hudson \& Moore 2000); (b) 0.8 MeV protons (Hudson \& Moore 2000) and (c) 46 MeV Ni ions (this work).} \label{fig:OCN-W33}
\end{figure}

Following Hudson et al. (2001) a possible route for the production of OCN$^-$, from the radiolysis of interstellar ice analogues, occurs via the reaction of NH or NH$_2$ radical with CO, leading to the formation of isocyanic acid molecule HNCO:

\begin{equation} \label{eq:nh3}
\textrm{NH}_3 \stackrel{CR}{\longrightarrow} (\textrm{NH}_3^+ + e^-)  \textrm{ or }   (\textrm{NH}_2 + \textrm{H})\\
\end{equation}
\begin{equation} \label{eq:nh4}
\textrm{NH}_3 + \textrm{NH}_3^+ \rightarrow \textrm{NH}_4^+ + \textrm{NH}_2 \\
\end{equation}
\begin{equation} \label{eq:nh4_b}
\textrm{NH}_2 + \textrm{CO}  \rightarrow \textrm{H} + \textrm{HNCO}  \quad \textrm{or} \quad  \textrm{NH} + \textrm{CO}  \rightarrow  \textrm{HNCO}  \\
\end{equation}

The reaction between isocyanic acid and ammonia (via acid-base reaction) produces ammonium, NH$_4^+$, and OCN$^-$:

\begin{equation} \label{eq:nh4_c}
\textrm{HNCO} + \textrm{NH}_3 \rightarrow \textrm{NH}_4^+ + \textrm{OCN}^-
\end{equation}

In the infrared spectra of the irradiated ices by heavy ions, beyond the OCN$^-$ band, the HNCO and NH$_4^+$ species are also observed at frequencies around 2265 and 1490 cm$^{-1}$, respectively.

A comparison between the infrared spectra of interstellar ices around the embedded protostar W33A with irradiated ammonia-water-CO mixtures from laboratory is given in Fig.~\ref{fig:OCN-W33}. The highest curve presents the IR spectrum recorded by Infrared Space Observatory (ISO) toward the protostellar object W33A. The data have been extracted from ISO database\footnote{http://iso.esac.esa.int/ida}.
Curve (a) shows the IR spectrum of H$_2$O:NH$_3$:CO (1:0.2:0.2) ice at 15 K processed 10 minutes of UV photons from hydrogen-discharge
lamp (Hudson \& Moore 2000). Curve (b) presents the IR spectrum of H$_2$O:NH$_3$:CO (1:0.2:0.2) ice at 15 K irradiated by 0.8 MeV protons at a dose of 19 eV molec$^{-1}$ (Hudson \& Moore 2000). The lowest curve (c) shows the infrared spectrum of H$_2$O:NH$_3$:CO (1:0.6:0.4) ice at 13 K observed after the bombardment by 46 MeV Ni ions at fluences around 2$\times 10^{13}$ ions cm$^{-2}$ (this work).
The radiolysis of H$_2$O:NH$_3$:CO performed by the present study, using heavy and energetic cosmic rays, reproduces very well the OCN$^-$ (2165 cm$^{-1}$) and CO (2139 cm$^{-1}$) infrared features observed in the W33A spectrum. Other spectral features like the broad IR peaks observed at 1650 and 1450 cm$^{-1}$ are also similar to the astronomical source.

\subsubsection{Ammonium, NH$^+_4$}

The broad $\nu_4$ vibration mode of NH$^+_4$ around 1490 cm$^{-1}$ is also observed after the heavy ion bombardment on both ammonia-containing ices (see Figs~\ref{fig:EXP1-FTIR}a-b). This line also has been observed from similar irradiation experiments involving photolysis and radiolysis of ammonia-containing ices (e.g. Demyk et al. 1998; Hudson et al. 2001; Hudson \& Moore 2000). As discussed by Demyk et al. (1998), the band strength of the NH$^+_4$ $\nu_4$ line is very uncertain (due to the broad profile in the astrophysical ice analogs). This fact, combined with a possible contamination from other infrared lines, makes this particular column density determination very imprecise.

The NH$^+_4$ formation directly from the processing of ammonia is given by Eqs.~\ref{eq:nh3} and \ref{eq:nh4}. The reaction pathways due to the processing of H$_2$O:NH$_3$:CO ice are given by Eqs.~\ref{eq:nh4_b} and \ref{eq:nh4_c}.  As discussed by Moore et al. (2007), the presence of water in interstellar ices also contributes to convert NH$_3$ into NH$_4^+$:
\begin{equation}
\textrm{H$_2$O} \stackrel{CR}{\longrightarrow} \textrm{H$_2$O}^+ + e^-
\end{equation}

\begin{equation}
\textrm{H$_2$O}^+ + \textrm{H$_2$O} \rightarrow \textrm{OH} + \textrm{H$_3$O}^+
\end{equation}

\begin{equation}
\textrm{H$_2$O}^+ + \textrm{NH}_3 \rightarrow \textrm{OH} + \textrm{NH}_4^+
\end{equation}

\begin{equation}
\textrm{H$_3$O}^+ + \textrm{NH}_3 \rightarrow \textrm{H$_2$O} + \textrm{NH}_4^+
\end{equation}

Besides the observation of the OCN$^-$ ion, the occurrence of ammonium at several protostellar disks was also observed (Gibb et al 2000; Boogert et al. 2004; Gibb et al. 2004; Demyk et al. 1998 and references therein) and possibilly for several solar system ices (Moore et al. 2007). Following Muñoz Caro \& Shutte (2003), $NH_4^+$ is one of the species involved in the formation of highly complex molecules like hexamethylenetetramine HMT, (CH$_2$)$_6$N$_4$, which was observed in experiments involving UV photoprocessing of interstellar ammonia-containing ice analogs.

\subsubsection{Other species}

\begin{figure}[!t]
 \centering
\includegraphics[scale=0.72]{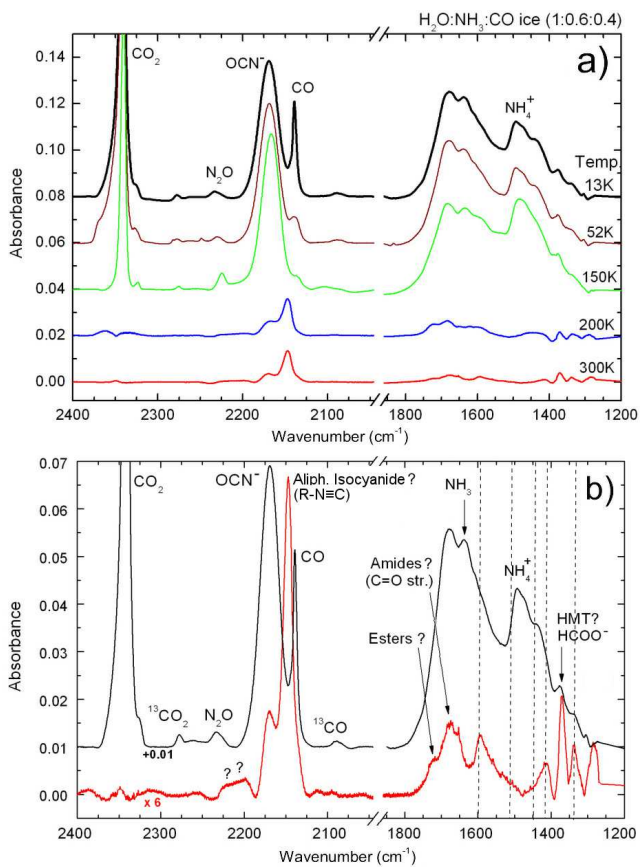}
\caption{(a) Infrared spectra of H$_2$O:NH$_3$:CO ice (1:0.6:0.4) from 2400 to 1200 cm$^{-1}$ during heating up to room temperature. The sample temperature for each spectrum is given. Each spectrum has an offset of 0.02 for better visualization. (b) Comparison between the irradiated ice at 13 K (top spectrum) and the 300 K residue (bottom spectrum). Vertical dashed lines indicate the frequencies of some vibration modes of zwitterionic glycine (NH$_3^+$CH$_2$COO$^-$).} \label{fig:heating}
\end{figure}

The infrared spectra of H$_2$O:NH$_3$:CO ice (1:0.6:0.4) from 2400 to 1200 cm$^{-1}$ acquired during the slow sample heating, from 13 K up to room temperature, are shown in Fig.~\ref{fig:heating}a. For each spectrum the sample temperature is indicated. The spectra have an offset of 0.02 for better visualization. The CO line at 2139 cm$^{-1}$ decreases drastically in a temperature between 13 and 52 K. At 150 K the N$_2$O peak ($\sim$ 2234 cm$^{-1}$) increases slightly. The unidentified feature around 1450 cm$^{-1}$ is not seen anymore above this temperature. At 200 K all the water, ammonia and CO$_2$ ices have been sublimated and their peaks have disappeared. Although very attenuated, the OCN$^-$ peak ($\sim$2169 cm$^{-1}$) is still present at 200 K. At this temperatures and higher, we observe the appearance of new  unidentified line around 2147 cm$^{-1}$ (between the OCN$^-$ and CO peaks). This peak is tentatively assigned to aliphatic isocyanide R-N$\equiv$C molecules (2150 cm$^{-1}$) as observed by Imanaka et al. (2004) in UV photolysis of N$_2$:CH$_4$ ices.

A comparison between the irradiated ice at 13 K and the organic residue at 300~K is shown at Fig.~\ref{fig:heating}b. Tentative band attribution is indicated.  At room temperature, a large structure (possibly two overlapping peaks) is observed around 2210 cm$^{-1}$. Since these features are close to C-N stretch vibration band, we suggest that they may be due to organic residues containing CN bearing molecules. Vertical dashed lines indicate the possible frequency of some vibration modes of the zwitterionic glycine (NH$_3^+$CH$_2$COO$^-$).

The observed infrared bands produced by the radiolysis of the H$_2$O:NH$_3$:CO ice (1:0.6:0.4) at 13 K by heavy, charged and energetic ions in the  2400 to 1200 cm$^{-1}$ range are listed in Table~\ref{tab:newspec}. The lines observed after the sample heating to room temperature are also shown. The assignments are proposed by comparing the spectra with those of similar experiments by photolysis or radiolysis of interstellar ice analogs (Hudson et al. 2001; Jamieson et al. 2005; Hudson \& Moore 2000; van Broekhuizen et al. 2005; Munoz Caro and Shutte 2003; Holtom et al. 2005; Demyk et al. 1998; Imanaka et al. 2004). The low frequency lines are tentatively assigned to N$_x$O$_y$ molecules (Jamieson et al. 2005).

\begin{table}[!t]
\begin{center}
\caption{Assignment of infrared absorption features produced by the radiolysis of the H$_2$O:NH$_3$:CO ice (1:0.6:0.4) by 46 MeV $^{58}$Ni$^{13+}$ ions at 13 K and after warming to 300 K.} \label{tab:newspec}
\setlength{\tabcolsep}{4pt}
\begin{tabular}{ l l l l r }
\hline \hline
Frequency  & Wavelength & Temp.  & Molecule                      & Notes \\
(cm$^{-1}$)& ($\mu$m) & (K)       &                              &       \\
\hline
2233           & 4.48 & 13       & N$_2$O                          & [1,2]  \\
2218-2200      & 4.51-4.54 & 300 &  nitriles$^\dag$                & [8]  \\
2168           & 4.61 & 13, 300  & OCN$^-$                         & [1,3,4,7] \\
2147           & 4.66 & 300      & aliph. isocyanide$^\dag$        & [8] \\
$\sim$ 2112    & 4.73 & 300      & NCO$_2^\dag$                    & [2] \\
1725           & 5.80 & 300    & ester$^\dag$                      & [5]  \\
1683           & 5.94 & 300    & amides$^\dag$                     & [5]  \\
1652           & 6.05 & 300    & asym-N$_2$O$_3^\dag$              & [2]  \\
1637           & 6.11 & 13     & ?                                 &      \\
1593           & 6.28 & 300    & NH$_3^+$CH$_2$COO$^{-\dag}$       & [6]  \\
1558           & 6.42 & 300    & ?                                 &      \\
1533           & 6.52 & 300    & ?                                 &       \\
1506           & 6.64 & 300    & NH$_3^+$CH$_2$COO$^{-\dag}$       &  [6]  \\
$\sim$1490     & 6.71 & 13     & NH$_4^+$                          & [1,3,7] \\
1474           & 6.78 & 13     & NO$_3^\dag$                           & [2]  \\
1440           & 6.94 & 13     & NH$_3^+$CH$_2$COO$^{-\dag}$       & [6]  \\
1415           & 7.07 & 300    & NH$_3^+$CH$_2$COO$^{-\dag}$       & [6]  \\
$\sim$ 1370    & 7.30 & 13, 300 & HMT$^\dag$                      & [5]  \\
               &      &         & HCOO$^{-}$                       & [5,9]  \\
$\sim$1338     & 7.47 & 13, 300 & NH$_3^+$CH$_2$COO$^{-\dag}$      & [6]  \\
               &      &         & NH$_2$CH$_2$COO$^{-\dag}$              & [6]  \\
               &      &         & HCOO$^-$                         & [9]  \\
1305           & 7.66 & 13      & N$_2$O$_3^\dag$; N$_2$O$_4^\dag$ & [2] \\
1283           & 7.80 & 300     & N$_2$O$^\dag$                    & [2] \\
\hline \hline
\end{tabular}
\end{center}
\vspace{-0.6cm}
 \begin{flushleft}
 $^\dag$Tentative assignment.
\end{flushleft}
\vspace{-0.3cm}
 [1] Proton bombardment of several ices (Hudson et al. 2001); [2] Electron bombardment of N$_2$:CO$_2$ (Jamieson et al. 2005); [3] Hudson \& Moore 2000; [4] van Broekhuizen et al. 2005; [5] UV photolysis of H$_2$O:NH$_3$:CH$_3$OH:CO:CO$_2$ ice (Munoz Caro and Shutte 2003); [6] Electron bombardment of CH$_3$NH$_2$:CO$_2$ ice (Holtom et al. 2005); [7] UV photolysis of ammonia-containing ices (Demyk et al. 1998); [8] UV photolysis of N$_2$:CH$_4$ ices at various pressures (Imanaka et al. 2004); [9] proton bombardment of H$_2$O:NH$_3$:CO ice (Hudson \& Moore 2001).
\end{table}

At room temperature, the peaks around 1593, 1506, 1415, 1338 cm$^{-1}$ may possibly be assigned to different vibration modes of zwitterionic glycine (NH$_3^+$CH$_2$COO$^-$). This amino acid was observed among the residues coming from 5 keV electron bombardment of CH$_3$NH$_2$:CO$_2$ (30:1) ice (Holtom et al. 2005) at 10 K. Zwiterionic glycine has been also observed after irradiation of NH$_3$:CH$_3$COOD (1:1) ice by low energy electrons (Lafosse et al. 2006) at 25 K. The authors have observed zwitterionic glycine even at 25 K, without any subsequent thermal activation.

The presence of amino acid precursors has been observed among the organic residues due to photolysis and radiolysis of several ammonia-containing interstellar ice analogs. The investigation employed vacuum ultraviolet photons from synchrotron light sources (Chen et al. 2008; Nuevo et al. 2007; Nuevo et al. 2006), hydrogen UV-lamp (Elsila et al. 2007; Muñoz-Caro and Schuttle 2003; Bernstein et al. 2002; Muñoz caro et al. 2002) and recently, by ion bombardment (Takano et al. 2007; Kobayashi et al. 2008).

The IR peak centered around 1370 cm$^{-1}$ observed in both 13 K and 300 K spectra was tentatively assigned to hexamethylenetetramine - HMT, (CH$_2$)$_6$N$_4$, since a similar feature was also observed in the UV photolysis of H$_2$O:NH$_3$:CH$_3$OH:CO:CO$_2$ ice by Muñoz Caro and Shutte (2003). The observation of NH$_4^+$ in the same organic residue reinforces the evidence for the presence of HMT, since the ammonium is an essential component for its formation.

The presence of these complex molecules in the organic residue from the bombardment of interstellar ice analogs by energetic and heavy ions has to be confirmed. However, the result of this experiment suggests that even deeply inside molecular cores or other denser regions like protoplanetary disks, the interstellar grains are being highly transformed. The produced organic molecules, trapped into and onto dust grains, meteoroids and comets, could be delivered into the planets/moons possibly allowing pre-biotic chemistry in such environments where water is also found in liquid state.

\subsection{Astrophysical implication}

\begin{figure*}[!t]
 \centering
\includegraphics[scale=0.21]{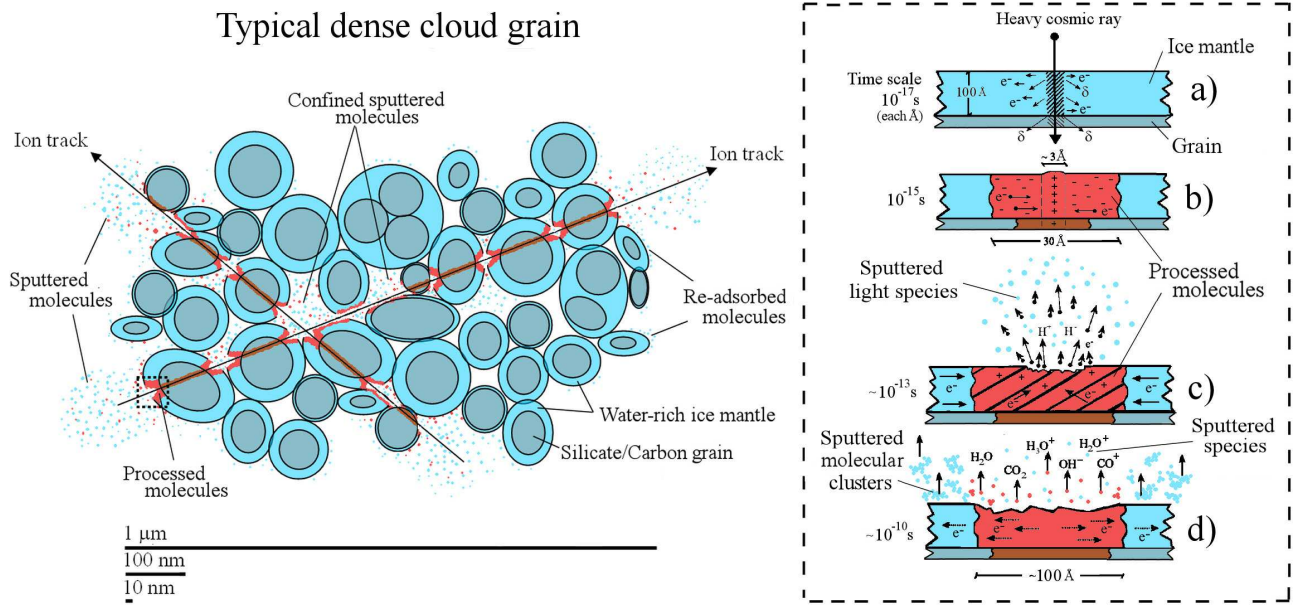}
\caption{Schematic view showing the interaction between heavy ion cosmic ray with a typical interstellar grain inside dense clouds. The ion track along the coagulate sub-micron size grains, the grain mantles, the processed and the sputtered molecules are indicated. Figure insets were adapted from Andrade et al. (2008) and indicate the physical-chemical changes on the grain mantle due to the impact with heavy ion. See details in text.} \label{fig:astrof}
\end{figure*}

Inside dense regions of interstellar medium, where direct stellar UV is extremely attenuated, X-rays (despite some attenuation degree) and cosmic rays are the main drivers for the gas phase and grain surface chemistry. For example, the penetration depth of 46 MeV Ni ions and equivelocity protons inside water ice is about 1 and 3 orders of magnitude higher than for 500 and 50 eV photons (considering the photoabsorption cross section from Chan et al. (1993) and McLaren et al. (1987), respectively). In these environments the dust grains can reach a size of the order of a micron, as a coagulation of sub-micron grains with ice-rich mantles of tens of nanometers (Mathis 1990, Mathis, Rumpl, Nordsieck 1977; Li \& Greenberg 1997; Chokshi, Tielens \& Hollenbach 1993). However, the presence of larger grains (radii within one to tens of microns) has also been suggested (Witt, Smith \& Dwek 2001; Taylor Baggaley \& Steel 1996; Gr\"{u}n et al. 1994).

Figure~\ref{fig:astrof} illustrates, as a schematic view, the interaction between heavy ion cosmic rays with a typical interstellar grain inside dense clouds. The ion track (along the coagulated sub-micron size grains), the grain mantles, the processed material and the sputtered molecules are indicated. The inset of the figure was adapted from Andrade et al. (2008) and shows the differents stages (times scales) of physico-chemical changes on the grain mantel caused by the heavy ion impact. At the pertinent heavy ion velocities, the projectile traverses atomic distances of the grain mantle in $10^{-17}$~s. The energetic cosmic ray analog ionizes and excites molecules along its trajectory inside the solid (inset fig.~\ref{fig:astrof}a). The  secondary electrons ($\delta$) move away from the projectile trajectory, generating a positive infratrack in the
center and a negative ultratrack around. Afterwards, molecular dissociations and chemical reactions occur, causing sputtering of light ions such as H$^+$ (inset fig.~\ref{fig:astrof}c). Following Iza et al. (2006) the sputtering occurs mainly due to long lived repulsive electronic excitation which lead to atomic or molecular motion.

After the Coulomb repulsion of ionizing species, the next stage is the track relaxation, when preformed or newly formed larger chemical species (molecular clusters) are desorbed (inset fig.~\ref{fig:astrof}d) at timescales around $\sim10^{-10}$~s. Shock waves transfer momentum to material at the ice surface which desorb as molecules from a site located as far as 3-5 nm of the ion track.

Due to the high porosity of grains inside dense clouds (coagulated from sub-micron heterogenic grains), some sputtered species inside the grain pores become confined, being re-adsorbed on the grains at a later time. The re-adsorption of sputtered species (processed or not) my also occur at the outer layer of the micron size grain mantle. The projectile energy deposited along the interfaces between the water-rich mantle and the silicate/carbon grain may promote chemical mixing within these two different chemical regions.

Inside dense interstellar regions,  the 1 MeV proton flux is around 1 proton cm$^{-2}$ s$^{-1}$ (Mennella et al. 2003; Morfill et al. 1976). Inside these regions the iron to proton ratio is about 1.6 $\times 10^{-4}$, and the iron fraction of the canonical cosmic ray flux is estimated to be  $2 \times 10^{-3}$ cm$^{-2}$ s$^{-1}$ (Roberts et al. 2007; Léger et al. 1985). Following Drury et al. (1999), heavy ions with 12 $\lesssim Z \lesssim$ 29 (e.g. from Mg to Cu) present roughly the same abundance, while those with Z $>$ 30 are more than 3 orders of magnitude less abundant. The average cosmic ray flux composed by heavy nuclei ions (Z $\gtrsim$ 12) is estimated to be $\phi_{HCR} \sim 5 \times 10^{-2}$ cm$^{-2}$ s$^{-1}$. Despite this small value, the sputtering promoted by heavy and energetic ions can be more than 4 orders of magnitude higher than ones promoted by the abundant low energy protons (Léger et al. (1985), Seperuelo Duarte et al. (2009a,b)). Moreover, in regions where the ionizing stellar photons are very attenuated, the sputtering promoted by cosmic rays becomes a competitive desorption process. This issue should be taken in consideration in future chemical models of inner regions of molecular clouds where a considerable fraction of excited molecules are released into the gas phase by heavy ion impact.

\begin{table}[!t]
\begin{center} %
\caption{Average dissociation rate ($R$) and the corresponding half life ($t_{1/2}$) of water, ammonia and CO molecules under bombardment by 46 MeV Ni ions; similar results are expected for heavy cosmic rays interaction with a interstellar ammonia-containing (H$_2$O:NH$_3$:CO (1:0.6:0.4)) grain.} \label{tab:halflife}
\setlength{\tabcolsep}{5pt}
\begin{tabular}{ l l l l }
\hline \hline
Species         & Relative initial  & R                       &  $t_{1/2}$      \\
                & abundance         & ($10^{-15}$ s$^{-1}$)   &  ($10^6$ years)   \\
\hline
H$_2$O          & 50\%              & 8    & 2      \\
NH$_3$          & 30\%              & 7    & 3      \\
CO              & 20\%              & 9    & 2     \\
\hline \hline
\end{tabular}
\end{center}
\end{table}

A rough estimation of half lives of molecular species in space can be carried out by considering only molecular dissociation effects of cosmic rays. The average dissociation rate, $R$, of a given species $i$, induced by the interstellar heavy cosmic ray field is approximatively given by:
\begin{equation} \label{eq-R}
R_i \approx \phi_{HCR} \times \sigma_{d,i} \quad \textrm{[s$^{-1}$]}
\end{equation}
where $\sigma_{d, i}$ is the average dissociation cross section of a frozen species $i$ due to heavy cosmic rays  and $\phi_{HCR}$ is the average heavy ion cosmic ray flux (ions cm$^{-2} s^{-1}$). The half-life, $t_{1/2}$, for a given species $i$ may be obtained from Eq.~\ref{eq-R} by writing:
\begin{equation} \label{eq-time}
t_{1/2,i}=\frac{\ln 2}{R_i} \quad \textrm{[s]}
\end{equation}
which does not depend on the molecular number density in the ice.
The dissociation cross section depends on the projectile energy: for Ni ions, $\sigma_{d, i}$ is typically 2$\times 10^{-13}$ cm$^{2}$ at 46 MeV, decreasing to 0.3 $\times 10^{-13}$ cm$^{2}$ at 537 MeV (Seperuelo Duarte et al. 2009b).  Nevertheless, assuming the values of  $\sigma_{d, i}$  for Ni ions at 46 MeV as the average value and $\phi_{HCR} \sim 5 \times 10^{-2}$ ions cm$^{-2}$ s$^{-1}$ as the heavy ion cosmic ray flux, R can be estimated.

Table~\ref{tab:halflife} presents the average dissociation rate and half life of water, ammonia and CO molecules bombarded by heavy ions  The estimated half life for the studied species over a typical heavy cosmic ray flux is about 2-3 $\times 10^6$ years. This value is in a good agreement with the half lives of typical molecular clouds and protostellar clouds (e.g. Millar \& Nejad 1985; Takeuchi et al. 2005). Better estimation for molecular dissociation rates and half lives promoted by the direct impact of heavy cosmic rays on interstellar ices depends on more accurate measurements/determination of the averaged heavy ion cosmic ray flux inside these dense environments.

\section{Summary and conclusions}

Experimental study on the interaction of heavy, highly charged and energetic ions (46 MeV $^{58}$Ni$^{13+}$) with ammonia-containing ices H$_2$O:NH$_3$ (1:0.5) and H$_2$O:NH$_3$:CO (1:0.6:0.4) is performed to simulate the physical chemistry induced by cosmic rays inside dense regions of interstellar medium like dense molecular clouds or protoplanetary disks. Our conclusions are:
\begin{enumerate}
  \item At the beginning of the irradiation, the ices become compact as revealed by the particular dependence of the column density on the fluence. Measuring the degree of compaction  through the integrated area of the OH dangling bond feature ($\sim$ 3650 cm$^{-1}$) from of water molecules trapped into ice micropores, the effect promoted by the impact of heavy ions seems to be at least 3 orders of magnitude higher than the one promoted by (0.8 MeV) protons.
  \item The infrared spectra of the irradiated samples present the production of several new species, including OCN$^-$, HNCO and NH$_4^+$. The OCN$^-$ band (2169 cm$^{-1}$) is observed even at room temperature. During the ice heating a broad feature was observed around 2218-2200 cm$^{-1}$ and was tentatively assigned to non-volatile nitriles. At room temperature an intense and sharp feature was observed at ($\sim$ 2150 cm$^{-1}$) and was tentatively assigned to aliphatic isocyanide ($\sim$ 2150 cm$^{-1}$).
  \item The spectra of the irradiated  H$_2$O:NH$_3$:CO ice (1:0.6:0.4) at room temperature reveal several bands which are tentatively assigned to zwitterionic glycine and HMT. Despite the presence of these complex molecules in the organic residue from the bombardments of interstellar ice analogs by energetic and heavy ions has to be confirmed.
  \item The obtained value for the dissociation cross section of water, ammonia and carbon monoxide (in the tertiary mixture ice) due to heavy cosmic ray analogs are $\sim 2 \times 10^{-13}$, 1.4$\times 10^{-13}$ and 1.9$\times 10^{-13}$ cm$^{2}$, respectively. These values seem not to be affected by small ($<$30\%) changes in the initial relative molecular abundances.
  \item In the presence of a typical heavy cosmic ray field the estimated half lives for the studied species is about 2-3 $\times 10^6$ years. This value is in a good agreement with the half lives of typical molecular clouds and protostellar clouds.
\end{enumerate}

Although they represent only a small fraction ($\sim$ 1\%) of the cosmic rays flux, some effects (e.g. molecular sputtering and ice compaction) promoted by heavy ions on interstellar ice grains are more intense than those promoted by protons. This should be taken under consideration in future chemical models of inner regions of molecular clouds since heavy ions will contribute molecules to the gas phase and trigger chemical surface and bulk reactions.

%
\begin{acknowledgements} The authors acknowledge the agencies COFECUB (France) as well as CAPES, CNPq and FAPERJ (Brazil) for partial support. We thanks  Th. Been, I. Monnet and Y. Ngono-Ravache for technical support. We also thank D. P. P. Andrade for fruitful discussions.
\end{acknowledgements}
%
%


\begin{thebibliography}{}

\bibitem{}
Andrade D.P.P., Boechat-Roberty H.M., da Silveira E.F., Pilling S., Iza P., Martinez R., Farenzena L.S., Homem M. G. P. \& Rocco M. L. M., 2008, J. Phys. Chem. C., 112, 11954

\bibitem{}
Bernstein M. P., Dworkin, J.P., Sandford S.A., et al. 2002, Nature, 416, 401

\bibitem{}
Bernstein M.P., Sandford S.A. \& Allamandola L.J., 2000, ApJ, 542, 894

\bibitem{}
Bird M.K., Janardhan P.,Wilson T.L., et al., 1999, Earth Moon Planets

\bibitem{}
Boogert A.A.C., et al, 2004, ApJS, 154, 359

\bibitem{}
Brown W.L., Lanzerotti L.J., Johnson R.E., 1982, Science, 218, 525

\bibitem{}
Brown W.L., Augustyniak W.M., Marcantonio K.J., Simmons E.H., Boring J.W., et al., 1984, Nucl. Instr. and Meth. B1, IV, 307

\bibitem{}
Chokshi A., Tielens A.G.G.M. \& Hollenbach D., 1993, ApJ, 407, 806

\bibitem{}
Collings M.P., Anderson M.A., Chen R., Dever J.W., Viti S., Williams D. A. \& McCoustra M.R. S., 2004, MNRAS, 354, 1133

\bibitem{}
Chan W.F., Cooper G. \& Brion C.E. 1993, Chem. Phys. 178, 387

\bibitem{}
Chiar J.E., Tielens A.G.G.M., Whittet D.C.B., Schutte W.A., et al. 2000, ApJ, 537, 749.

\bibitem{}
Chiar J.E., Adamson A.J., Pendleton Y.J., et al. 2002, ApJ, 570, 198

\bibitem{}
Chen Y.-J., Nuevp M., Yih T.-S, IP W.-H, Fung H.-S, Cheng C.-Y., Tsai H.-R. \& Wu C.-Y.R., 2008, MNRAS, 384, 605

\bibitem{}
Demyk K., Dartois E., d´Hendecourt L., Jourdain de Muizon M., Heras A. M. \& Breitfellner 1998, A\&A, 339, 553

\bibitem{}
Drury L.O.C., Meyer J.-P., Ellison D.C., 1999, ``Topics in Cosmic-Ray Astrophysics", M.A. DuVernois ed., Nova Science Publishers, New-York

\bibitem{}
Ehrenfreund P. \& Charnley S.B. 2000, ARAA, 38, 427

\bibitem{}
Ehrenfreund P. \& Shuttle W.A., 2000, Proceedings of IAU Symp. 197 - From molecular clouds to planetary system.

\bibitem{}
Elsila J.E., Dworkin J.P., Bernstein M.P. Martin M.P. \& Sandford S.A., 2007, ApJ, 660, 911

\bibitem{}
Famá M., Shi J. \& Baragiola R.A., 2008, Surface Science, 602, 156

\bibitem{}
Gerakines P.A., Scutte W.A., Greenberg J.M., van Dishoeck E. F., 1995, A\&A, 296, 810

\bibitem{}
Gibb E.L., Whittet D.C.B., Schutte W.A., et al., 2000, ApJ, 536, 347

\bibitem{}
Gibb E.L., Whittet D.C.B., Boogert A.C.A. \& Tielenes A.G.G.M., 2004, ApJSS, 151, 35

\bibitem{}
Gibb E.L., Whittet D.C.B. \& Chiar J.E., 2001, ApJ, 558, 702


\bibitem{}
Gomis O., Leto G., Strazulla G., 2004a, A\&A, 420, 405

\bibitem{}
Gomis O., Satorre M.A., Strazulla G., Leto G., 2004b, Plan. Spac. Sci. 52, 371

\bibitem{}
Grim R.J.A. \& Greenberg J.M., 1987, ApJ, 321, L91

\bibitem{}
Gr\"{u}n E., Gustafson B., Mann .I, Baguhl M., Morfill G.E., Saubach P., Taylor A. \& Zook H.A. 1994, A\&A, 286, 915

\bibitem{}
d´Hendecourt L.B., Alamandola L.J., Greenberg J.M., 1985, A\&A, 152, 130

\bibitem{}
d´Hendecourt L.B. \& Allamandola L.J., 1986, A\&ASS, 64, 453

\bibitem{}
Holtom P.D., Bennett C.J., Osamura Y., Mason N.J. \& Kaiser R.I., 2005, ApJ, 626, 940

\bibitem{}
Hudson R.L. \& Moore M.H. 2000, A\&A, 357, 787

\bibitem{}
Hudson R.L., Moore M.H. \& Gerakines P.A., 2001, ApJ, 550, 1140

\bibitem{}
Hudson R.L. \& Moore M.H., 2002, ApJ 568, 1095

\bibitem{}
Imanaka H., Khare B.N., Elsila J.E., Bakes E.L.O., McKay C.P., Cruikshank D.P., Sugita S., Matsui T. \& Zare R.N., 2004, Icarus, 168, 344

\bibitem{}
Iza P., Farenzena  L.S., Jalowy T., Groeneveld K.O. \& E.F. da Silveira, 2006, Nuclear Instruments and Methods in Physics Research Section B, 245, 61

\bibitem{}
Jamieson C.S., Bennett C.J.,Mebel A.M. \& Kaiser R. I., 2005, ApJ, 624, 436

\bibitem{}
Jenniskens P., Blake D.F. \& Kouchi A. 1998, Solar System Ices (Astrophysics and Space Science Library), Kluwer Academic Press, pg.139-156, Edts. B. Schmitt, C. de Bergh, M. Festou.

\bibitem{}
Keane J.V., Tielens A.G.G.M., Boogert A.C.A., Schutte W.A. \& Whittet D.C.B., 2001, A\&A, 376, 254

\bibitem{}
Kawakita et al. 2006, ApJ, 643, 1337

\bibitem{}
Kerkhof O., Schutte W.A., \& Ehrenfreund P. 1999, A\&A, 346, 990

\bibitem{}
Kobayashi K. ,Kaneko T., Takano Y. \&  Takahashi J.-I., 2008, Proc. IAU sympos. 251, pgs 465-472

\bibitem{}
Lacy J.H., Baas F., Allamandola L. J., et al. 1984, ApJ, 276, 533

\bibitem{}
Lacy J.H., Faraji H., Sandford S.A., Allamandola L.J., 1998, ApJ, 501, L105

\bibitem{}
Lafosse A., Bertin M., Domaracka A., Pliszka D., Illenberger E. \& Azria R., 2006, Phys. Chem. Chem. Phys., 8, 5564

\bibitem{}
Léger A., Jura M., Omont A., 1985, A\&A, 144, 147

\bibitem{}
Li A. \& Greenberg J.M., 1997, A\&A, 323

\bibitem{}
Loeffler M.J., Raut U., Vidal R.A., Baragiola R.A., Carlson R.W., 2006, Icarus, 180, 265

\bibitem{}
Mathis J.S., 1990, Annu. Rev. Astron. Astrophys., 28, 37

\bibitem{}
Mathis J.S., Rumpl W. \& Nordsieck K.H., 1977, ApJ, 217, 105

\bibitem{}
McLaren R., Ishii I., Hitchcock A.P. \& Robin M.B., 1987, J. Chem. Phys., 87, 4344

\bibitem{}
Moore M. H., Donn B., Khanna R. \& A´Hearn M.F., 1983, et al. 1983; Icarus, 54, 388

\bibitem{}
Moore M.H., Ferrante R.F., Hudson R.L., Stonee J.N., 2007, Icarus, 190, 260

\bibitem{}
Mennella V., Baratta G.A., Esposito A., Ferini G. \& Pendleton Y. J., 2003, ApJ, 587, 727

\bibitem{}
Millar T.J. \& Nejad L.A.M., 1985, MNRAS, 217, 507

\bibitem{}
Morfill G.E., Volk H.J., Lee M.A., 1976, J. Geophys. Res., 81, 5841

\bibitem{}
Muñoz Caro G.M. \& Shutte S.A., 2003, A\&A, 412, 121

\bibitem{}
Muñoz Caro G.M., et al., 2002, Nature, 416, 403

\bibitem{}
Nastasi M., Mayer J. \& Hirvonen J.K., 1996, in ``Ion-Solid Interactions: Fundamentals and Applications", Cambridge Solid State Science Series, Cambridge University Press

\bibitem{}
Novozamsky J.H., Schutte W.A. \& Keane J.V., 2001, A\&A, 379, 588

\bibitem{}
Nuevo M., Meierhenrich U.J., Muñoz Caro G.M., Dartois E., D'Hendecourt L., et al., 2006, A\&A, 457, 741	

\bibitem{}
Nuevo M., et al., 2007, Adv. Space Res., 39, 400

\bibitem{}
Palumbo M.E., Strazulla G., Pendleton Y.J., \& Tielens A.G.G.M. 2000, ApJ, 534, 801

\bibitem{}
Palumbo M. E., 2006, A\&A, 453, 903

\bibitem{}
Pontoppidan K.M., Dartois E., van Dishoeck, E.F., Thi W.-F. \& d´Hedencourt L., 2003, A\&A, 404, L17

\bibitem{}
Prasad S.S. \& Tarafdar S.P., 1983, ApJ, 267, 603

\bibitem{}
Pendleton Y.J., Tielens A.G.G.M., Tokunaga A.T. \& Bernstein M.P., 1999, ApJ, 475, 144

\bibitem{}
Roberts J.F., Rawlings J.M.C., Viti S. \& Williams D. A., 2007, MNRAS, 382, 733

\bibitem{}
Rowland B. \& Devlin J.P., 1991, J. Chem. Phys, 94, 812

\bibitem{}
Rowland B., Fisher M. \& Devlin J.P., 1991, J. Chem. Phys, 95, 1378

\bibitem{}
Raunier S., Chiavassa T., Marinelli F., Allouche A., \& Aycard J. P., 2003, J. Phys. Chem. A, 107, 9335

\bibitem{}
Schutte W.A. \& Greenberg J.M., 1997, A\&A, 317, L43

\bibitem{}
Shen C.J., Greenberg J.M., Schute W.A., \& van Dishoeck, E.F., 2004, A\&A, 415, 203

\bibitem{}
Seperuelo Duarte E., Boduch P., Rothard H., Been T., Dartois E., et al., 2009a, A\&A, 502, 599

\bibitem{}
Seperuelo Duarte E., Boduch P., Rothard H., Been T., Dartois E., et al., 2009b, submitted to A\&A.

\bibitem{}
Schmidt R., Schoppmann C., Brandle D., et al. 1991,  Phys. Rev. B, 44, 2

\bibitem{}
Soifer B.T., Puetter R.C., Russell R.W., Willner S.P., Harvey P.M., Gillett F.C., 1979, ApJ, 232L, 53

\bibitem{}
Spoon H. W. W., Moorwood A. F. M., Pontoppidan K. M., Cami J., Kregel M., et al., 2003, A\&A, 402, 499

\bibitem{}
Takano Y., Takahashi J.-I., Kaneko T., Marumo K., Kobayashi K., 2007, Earht and Planetary Science Letters, 254, 106


\bibitem{}
Takeuchi T., Clarke C.J. \& Lin D.N.C., 2005, ApJ, 627, 286

\bibitem{}
Taylor A.D., Baggaley W. J. \& Steel D.I.. 1996, Nature, 380, 323

\bibitem{}
Tegler S.C., Weintraub D.A., Allamandola L.J., et al., 1993, ApJ, 411, 260

\bibitem{}
Tielens A.G.G.M. \& Hagen W., 1982, A\&A, 114, 245

\bibitem{}
van Broekhuizen F.A., Keane J.V. \& Schutte W.A., 2004, A\&A, 415, 425

\bibitem{}
van Broekhuizen F.A., Pontoppidan K.M., Fraser H.J. \& van Dishoeck E.F., 2005, A\&A, 441, 249

\bibitem{}
Wang, F., Larkins, F. P., Brunger, M. J., Michalewicz, M. T., \& Winkler, D. A.
2001, Spectrochimica Acta A, 57, 9

\bibitem{}
Witt A. N., Smith R. K. \& Dwek E., 2001, ApJ, 550, L201

\bibitem{}
Whittet D.C.B., Pendleton Y. J., Gibb E. L., Boogert A.C.A, Chiar J.E. and Nummelin A., ApJ, 2001, 550, 793

\bibitem{}
Ziegler J. F., \& Biersack, J. P. (2006), www.srim.org, version 2006.02
\end{thebibliography}
\end{document}